\documentclass[twocolumn,prb,showpacs]{revtex4}
\usepackage{graphicx}

\usepackage{dcolumn}

\usepackage{tabularx}

\usepackage{float}

\usepackage{amsmath}

\begin{document}


\title{Optical conductivity from local anharmonic phonons}

\author{Hideki Matsumoto}%
\email{matumoto@ldp.phys.tohoku.ac.jp}
\affiliation{
Department of Physics, Graduate School of Science, Tohoku University, Sendai 
980-8578 Japan} 
\affiliation{%
Institute for Materials Research, Tohoku University, Sendai, 980-8577 Japan 
}%
\affiliation{ 
CREST(JST), 4-1-8 Honcho, Kawaguchi, 
Saitama 332-0012, Japan}
\author{Tatsuya Mori}
\affiliation{
Department of Physics, Graduate School of Science, Tohoku University, Sendai 
980-8578 Japan} 
\author{Kei Iwamoto}
\affiliation{
Department of Physics, Graduate School of Science, Tohoku University, Sendai 
980-8578 Japan} 
\author{Shohei Goshima}
\affiliation{
Department of Physics, Graduate School of Science, Tohoku University, Sendai 
980-8578 Japan} 
\author{Syunsuke Kushibiki}
\affiliation{
Department of Physics, Graduate School of Science, Tohoku University, Sendai 
980-8578 Japan} 
\author{Naoki Toyota}
\email{toyota-n@ldp.phys.tohoku.ac.jp}
\affiliation{
Department of Physics, Graduate School of Science, Tohoku University, Sendai 
980-8578 Japan}

\date{February 12, 2009, revised May 19, 2009}

\begin{abstract}
	Recently there has been paid much attention to phenomena caused 
by local anharmonic vibrations of the guest ions encapsulated in 
polyhedral cages of materials such as pyrochlore oxides, filled
skutterdites and clathrates. We theoretically investigate the optical 
conductivity solely due to these so-called rattling phonons in a 
one-dimensional anharmonic potential model. The dipole interaction of the
guest ions with electric fields induces excitations expressed as 
transitions among vibrational states with non-equally spaced energies, 
resulting in a natural line broadening and a shift of the peak 
frequency as anharmonic effects. 
In the case of a single well potential, a softening of the peak 
frequency and an asymmetric narrowing of the line width with decreasing 
temperature are understood as a shift of the spectral weight to lower 
level transitions. On the other hand, the case of a double minima 
potential leads to a multi-splitting of a spectral peak in the 
conductivity spectrum with decreasing temperature.  
\end{abstract}

\maketitle
\def\pdt{\frac{\partial}{\partial t}}
\def\dx{\nabla_x}
\def\dy{\nabla_y}
\def\dz{\nabla_z}
\def\hf{\frac{1}{2}}

\makeatletter
\renewcommand{\theequation}{%
	\arabic{section}.\arabic{equation}}\@addtoreset{equation}{section}
\makeatother

\section{Introduction}
Anharmonicity in lattice vibrations has been one of the old problems
in condensed-matter physics\cite{Born1968,Madelung1978}.
Anharmonic effects in acoustic phonons were treated by 
perturbation theory\cite{Choquard1967,Madelung1978}, while those 
in local vibrations were investigated in impurities or disordered 
systems \cite{Stevenson1966,Newman1969,Barker1975}. 
It was pointed out that the effects of the anharmonicity 
in local vibrations appear in the characteristic 
temperature dependence of the vibrational 
frequency and of the line width\cite{Elliot1965}. 
Since an isolated irregular atom receives an anharmonic potential 
from the regular lattice, analysis have been made 
mostly on the anharmonic oscillation receiving effects of 
surrounding oscillation of the regular lattice. 

Recently a revised interest on anharmonic phonons has arisen 
in relation to a material series of pyrochlore oxides
\cite{Subramanian1983,Hanawa2001}, 
filled skutterdites\cite{Jeitschko1977,Braun1980,Sales1997} 
and clathrates\cite{Nolas1998,Sales2001,Bentien2004,Avila2006,Avila2008}.
Those materials, which are usually electrical conductors like metals, 
semimetals or heavily-doped semiconductors, have a common feature 
that some numbers of atoms form a three-dimensional network of polyhedral cages, 
in each of which a guest ions is accommodated. When the cages are oversized, 
the guest ion vibrates with a large amplitude in an anharmonic potential. 
Such vibrations are named as {\it the rattling phonons}. To note, depending on
kinds of guest ions, on-centering or off-centering vibrations
occur even in the same cage structure. 

	There have been reported various anomalous phenomena in the above materials, 
some of which have been discussed in relation to those rattling 
phonons\cite{reviews}. In applying some clathrate compounds to thermoelectric 
material devices, for example, the rattling phonons, in particular off-centered, 
are expected to suppress strongly  
the thermal conductivity by effectively scattering Debye-like acoustic phonons 
propagating through the cage network and carrying heat entropy
\cite{Bentien2004, Avila2006, Avila2008, Nakayama2008}. 
An alternative example is found in the superconductivity in a $\beta$-pyrochlore 
oxide KOs$_2$O$_6$. It was suggested that rattling vibrations of the K$^+$ in the 
OsO$_6$ octahedral cage were responsible for the strong-coupling superconductivity 
and also for an electron-mass enhancement\cite{Bruhwiler2006, Hiroi2007}. 
It may be fair, however, to state that these interesting issues as for the 
question how rattling phonons interact with cage acoustic phonons and/or charge 
carriers are far from being well understood\cite{reviews}. 

So far lattice vibrational modes including rattling phonons in 
the above cage materials have been studied rather extensively with use of 
spectroscopic measurements such as inelastic neutron and Raman scatterings\cite{reviews}. 
Some low-lying rattling modes are clarified to exhibit softening with decreasing 
temperatures. The softening phenomenon has been well recognized as one of the 
anharmonic effects from rattling phonons, which was discussed with a quasi-harmonic 
approximation\cite{Dahm2007}. Beside these spectroscopies, an infrared-active optical 
measurements, particularly in the Terahertz range, would be, in principle,
a powerful tool to clarify the charge dynamics in low-lying optical phonons near 
$q\sim 0$ with available optical conductivity spectra.

Recently, time-domain terahertz 
spectroscopy\cite{Mori2008a} has been successfully 
applied for the first time to observe the rattling phonons 
around 1THz in a type-I clathrate Ba$_8$Ga$_{16}$Ge$_{30}$ (BGG) \cite{Mori2008b}. 
In this paper, we investigate systematically 
the optical conductivity spectra from the rattling phonons 
in an on-centered and off-centered potential, and apply 
the obtained theoretical classification of the spectra to analyse the experimental  
result of BGG. The detail of the experiment 
will be presented elsewhere\cite{Mori2008b}. 
We express the optical conductivity by level transitions among 
states in an anharmonic potential\cite{Foster1993}. 
The present analysis shows that the 
most important effect of an anharmonic potential is  
the non-equal energy spacing of level transitions, resulting in 
an intrinsic and asymmetric spread of the excitation 
spectral, in contrast to the harmonic case of the equal 
energy spacing simply resulting in the Lorentzian spectral shape. 
Also there arise variety of level schemes for low-lying states depending 
on an on-centered or off-centered potential. 
Those features induce the natural broadening of 
the line width at higher temperature, softening of the peak 
frequency\cite{note1} and multi-peak structures at low 
temperature in optical conductivity. 

In the present paper, a one-dimensional model 
is used for simplicity. Essential effects from the anharmonicity 
are included. In the next section, the model and the expression 
of the optical conductivity are presented. In Sect.3 some of 
numerical results are presented. 
Comparison with the recent experiment\cite{Mori2008b} 
is discussed. Sect.4 is devoted to the conclusion. 
\section{Formulation}
As a model to describe the motion of a guest ion in the cage, 
we take the following one-dimensional anharmonic potential model, 
for simplicity, 
\begin{equation}
	H=\frac{p^2}{2M}+\frac{1}{2}kx^2+\frac{1}{4}\lambda x^4\ , 
	\label{H}
\end{equation}
where $M$, $p$ and $x$ are the mass, momentum and 
spatial coordinate of the guest ion, respectively. 
We neglect effects of acoustic phonons and electrons. 

The optical conductivity from the guest ion is obtained 
by considering the polarization induced by an applied 
oscillatory electric field $E(t)$, 
\begin{equation}
	H_I=-qE(t)x\ ,
\end{equation}
where $q$ is the charge of the guest ion. By use of the 
linear response theory, the polarization $P(t) (=\langle qx(t)\rangle)$ 
is obtained as
\begin{equation}
	P(t)=\frac{i}{\hbar}\int dt'q^2\langle Rx(t)x(t')\rangle E(t')\ ,
\end{equation}
where $\langle \cdots\rangle$ indicates the thermal average and 
$"R"$ means the retarded function. By taking the Fourier transform, 
the polarizability $\alpha(\omega)$ defined by
\begin{equation}
	P(\omega)=\alpha(\omega)E(\omega)
\end{equation}
is obtained as
\begin{equation}
	\alpha(\omega)=-\frac{q^2N}{\hbar}G_{xx}(\omega)\ ,
\end{equation}
where the density of the quest ion $N$  is taken into account and 
$G_{xx}(\omega)$ is defined by 
\begin{equation}
	\langle R x(t)x(t')\rangle
	=\frac{i}{2\pi}\int d\omega e^{-i\omega(t-t')}G_{xx}(\omega)\ .
\end{equation}
Let us denote eigenstates and eigenvalues of the Hamiltonian $H$ 
by $|n\rangle$ and $E_n$, respectively,
\begin{equation}
	H|n\rangle =E_n|n\rangle \ .
\end{equation}
Then $G_{xx}(\omega)$ is expressed as
\begin{equation}
	G_{xx}(\omega)
	=\sum_{nm}|\langle n|x|m\rangle|^2
		\frac{(e^{-\beta E_m}-e^{-\beta E_n})/Z}{
			\omega+i\Gamma_0/2-(E_n-E_m)/\hbar}
\end{equation}
with $\beta=1/k_BT$ and $Z=\sum_ne^{-\beta E_n}$. 
Here we have introduced a phenomenological 
parameter of the decay width $\Gamma_0$.  
Since the optical conductivity and the polarizability 
are related to each other as
\begin{equation}
	\sigma(\omega)=-i\omega\alpha(\omega)\ ,
\end{equation}
we have the complex optical conductivity as
\begin{eqnarray}
	\lefteqn{\sigma(\omega)/\sigma_0}\nonumber\\  
	&=&i\omega
			\sum_{\omega_{nm}>0}|\langle n|(x/x_0)|m\rangle|^2
			\frac{e^{-\beta E_m}-e^{-\beta E_n}}{Z}\nonumber\\ 
	&\times&\biggl(\frac{1}{\omega-\omega_{nm}+i\Gamma_0/2}
			-\frac{1}{\omega+\omega_{nm}+i\Gamma_0/2}\biggr)\ \label{sigEq}
\end{eqnarray}
with
\begin{equation}
	\omega_{nm}=(E_n-E_m)/\hbar\ .
\end{equation}
Hereafter, we use the notation $E_{nm}$, $\omega_{nm}$ and 
$\nu_{nm}$, respectively, as the energy, angular 
frequency and frequency for the transition from the $m$-state 
to the $n$-state. The normalization for the conductivity $\sigma_0$ 
is given by
\begin{equation}
	\sigma_0=q^2Nx_0^2/\hbar\ .
\end{equation}
with a length scale $x_0$ determined shortly.  
In the following analysis, we choose a suitable energy scale $\hbar\omega_0$, 
which can be, for example, the energy corresponding to 
1THz or a harmonic frequency evaluated from the quadratic term 
in the potential, or an observed phonon energy. The scaled parameters 
are defined as 
\begin{equation}
	\bar k=\frac{k}{M\omega_0^2}\ ,\ 
	\bar \lambda=\frac{\hbar\lambda}{M^2\omega_0^3}\ .\label{barlambda}
\end{equation}
and the length scale $x_0$ is given as
\begin{equation}
	x_0=\sqrt{\frac{\hbar}{M\omega_0}}\ .
\end{equation}

In the next section, we will calculate $\sigma(\omega)$ for  
cases of $k>0$, $k=0$ and $k<0$, and discuss characteristic 
features of the optical conductivity 
from the rattling phonon. 
\section{Numerical Analysis}
\subsection{The case of $k>0$}
We can choose $\bar k=1$ without loss of generality, 
In this case, $\omega_0$ is given by $\omega_0=\sqrt{k/M}$, 
the harmonic frequency. 
Let us estimate the magnitude of $\bar \lambda$. In the band 
structure calculation for Ba$_{8}$Ga$_{16}$Ge$_{30}$\cite{Madsen2005},
the coefficients $k$ and $\lambda$ were evaluated as 
$ka_B^2/2=2.813$mRy and $\lambda a_B^4/4=1.602$mRy, where 
$a_B$ is the Bohr radius. Then using the mass of the Ba-atom 
137.35 a.u., we have the harmonic frequency $\nu_0=0.6975$THz 
and $\bar\lambda=4.292\times 10^{-2}$. The actually 
observed phonon frequency lowest lying in Ba$_{8}$Ga$_{16}$Ge$_{30}$ 
is about 1THz and the parameters $k$ may vary about twice or so. 
Taking into account the above estimation and 
Eq.\ (\ref{barlambda}), we take the value of  
$\bar\lambda$ in the range of $\bar\lambda=0.0-0.05$. 

\begin{figure}[h]
\includegraphics[width=6truecm]{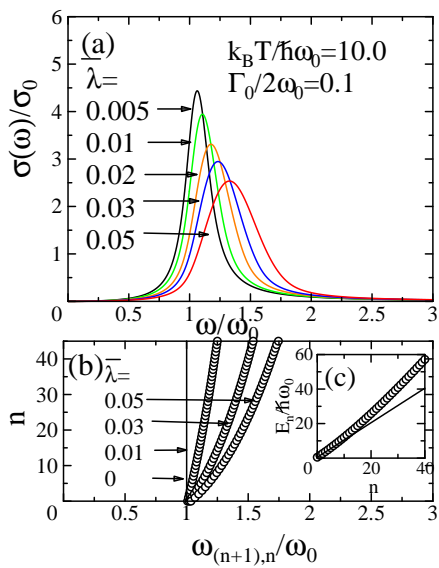}
\caption{(Color online) (a) Optical conductivity for various $\bar\lambda$ with $\bar k=1$ 
at $k_BT/\hbar\omega_0=10.0$, 
(b) excitation energy $\omega_{10}$   
and (c)energy eigenvalues for $\bar\lambda=0.05$. }
\label{F1_sigma10} 
\end{figure}
In Fig. \ref{F1_sigma10} (a) we plot the optical conductivity 
for various $\bar\lambda$ at $k_BT/\hbar\omega_0=10.0$. 
Hereafter, we introduce a constant decay width 
$\Gamma_0/2\omega_0=0.1$ by hand, in order to smooth the 
frequency dependence of the optical conductivity. How each level 
transition has a decay width depends on interactions 
with electrons or acoustic phonons. Details of such effects 
are not considered in this paper. Also in the following, $\sigma(\omega)$
denotes the real part of the complex conductivity.
At $\bar\lambda=0$, the phonon mode has the single 
angular frequency $\omega=\omega_0$, and the line width is 
the decay width $\Gamma_0$. As $\bar\lambda$ increases, 
both of the peak position and the line width increase. That is,  
the line width at higher temperature is determined by 
the anharmonic parameter $\bar\lambda$. In Fig. \ref{F1_sigma10} (b), 
we plot the excitation energy $\omega_{(n+1),n}=(E_{n+1}-E_n)/\hbar$. 
This is obtained by calculating the energy eigenvalues numerically. 
One example of the behavior of energy eigenvalues is shown in 
Fig. \ref{F1_sigma10}(c) for $\bar\lambda=0.05$. The label "n" is 
identical with the boson number in the harmonic case of $\bar\lambda=0$. 
The energy eigenvalues deviate from the linear behavior as 
$E_n\approx e_0+e_1n+e_2n^2$ but a perturbation calculation is not 
applicable, since $\bar\lambda n^2$ becomes an order of one for $n\sim 10$. 
The calculation shows that the transition probability arises mostly 
from $\langle n+1|x|n\rangle$. As is seen in Fig. \ref{F1_sigma10}(b), 
the excitation energies increase as eigenvalues(i.e. $n$) 
increase due to the anharmonicity; the larger the $\bar\lambda$, the more 
spreading excitation energies become. The non-equal energy spacing of 
the phononic level transitions leads to the 
intrinsic spread of the line width in the optical conductivity,  
as is seen in Fig. \ref{F1_sigma10}(a). It should be noted that 
excitations with larger $n$ ($\sim 30$) contribute at $k_BT/\hbar\omega_0=10$. 

\begin{figure}[h]
\includegraphics[width=6truecm]{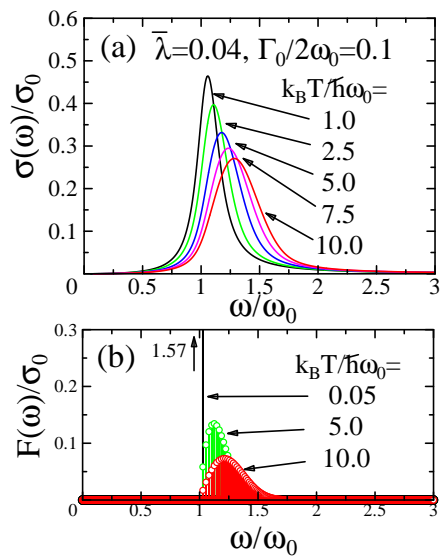}
\caption{(Color on line)(a)Temperature dependence of the optical conductivity 
	for $\bar\lambda=0.04$ and $\bar k=1$. (b)
The corresponding spectral weight at $k_BT/\hbar\omega_0$=0.05,5.0 and 10.0.} 
\label{F2_sigT} 
\end{figure}
In Fig. \ref{F2_sigT}(a), we plot the temperature dependence of 
the optical conductivity. The optical conductivity shows the 
softening of the peak position and the narrowing of the line width 
with decreasing temperature. 
At low temperature, only the first level 
transition contributes. As temperature increases, 
the transition involves higher energy levels, resulting 
in a shift of the effective peak position. 
Also the line width increases, since more level transitions 
contribute. In order to see the line-distribution  without 
the smoothening by the width $\Gamma_0$, we plot,   
in Fig. \ref{F2_sigT}(b), the spectral weight in the optical 
conductivity given by 
\begin{equation}
	F(\omega)/\sigma_0=\pi\sum_{\omega_{nm}=\omega}
	|\langle n|\xi|m\rangle |^2\ 
		\frac{e^{-\beta E_m}-e^{-\beta E_n}}{Z}\label{F}
\end{equation}
for $k_BT/\hbar\omega_0=0.05,5.0$ and 
$10.0$. Sharp lines arising from the non-equal spacing of the 
level transitions distribute rather densely in the present 
parameter values. Because of the Boltzmann factor and a factor 
$n$ from $|\langle n|\xi|n\pm 1\rangle |^2$, 
the maximum position of the weight increases with temperature and 
decreases exponentially as the energy $\hbar\omega$ increases. 
At $T=0$K, the main contribution arises from the transition 
between the ground state and the first excited state. 
The softening of the peak frequency and 
the sharpening of the line width with decreasing temperature are 
direct consequences derived solely from 
an inequivalence of the level spacing that is essential 
in the anharmonicity. Therefore measurements on the 
temperature dependence of the optical conductivity can 
provide a direct evidence for the anharmonicity.

\begin{figure}[h]
\includegraphics[width=6truecm]{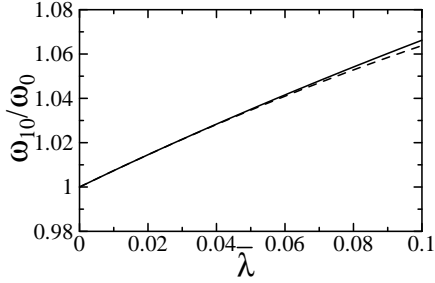}
\caption{$\bar\lambda$-dependence of $\omega_{10}$ for $\bar k=1$} 
\label{F3_nu10} 
\end{figure}
In Fig. \ref{F3_nu10} the transition energy for the 
$0\rightarrow 1$ transition is shown, which gives the 
peak frequency at $T=0$K. 
In the present parameter region, the curve is fitted by 
\begin{equation}
	\omega_{10}/\omega_0=1+0.7414\bar\lambda-0.7998{\bar\lambda}^2\ ,
\end{equation}
which deviates above $\bar\lambda\sim 0.05$ from the result of the perturbation, 
$\omega_{10}=1+(3/4)\bar\lambda-(9/8){\bar\lambda}^2$ (dashed line).  

\begin{figure}[h]
\includegraphics[width=6truecm]{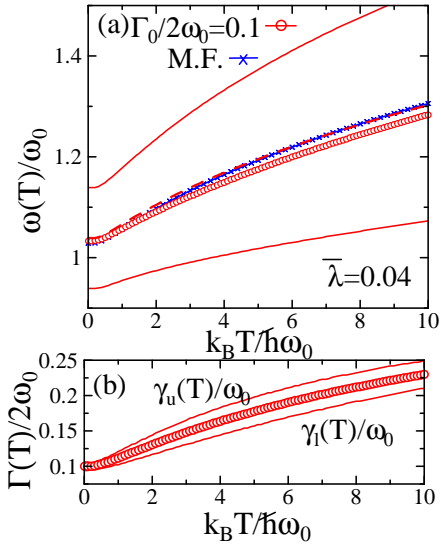}
\caption{(Color online) Temperature dependence of (a) the peak position and 
(b) line width for 
$\bar\lambda=0.04$ and $\bar k=1$} 
\label{F4_Peak_Gamma} 
\end{figure}
In Fig. \ref{F4_Peak_Gamma} (a) we plot the temperature dependence of 
the peak frequency for $\Gamma_0/2\omega_0=0.1$ (red circles) and,  
by red solid lines, upper and lower energies which give half 
values of the peak intensities. The red dashed line is for 
their averaged energy (say, mid-frequency). The blue diagonal-crosses indicate 
the result obtained from the following equation in the mean field theory\cite{Dahm2007} 
(The factor 3 is corrected.), 
\begin{equation}
	\frac{\omega^2}{\omega_0^2}
	=1+3\bar\lambda\frac{\omega_0}{\omega}\left(\frac{1}{
			\displaystyle{\exp\left(\frac{\hbar\omega_0}{k_BT}
			\frac{\omega}{\omega_0}\right)-1}}
				+\frac{1}{2}\right)\ .
\end{equation}
We see that 
the averaged value (red dashed line) well agree with 
the mean field results (blue diagonal-crosses).   
However, our result of the peak structure is asymmetric, 
reflecting the spectral weight, Eq.\ (\ref{F}).   
We express the line width at temperature $T$ as
\begin{equation}
	\Gamma(T)=\gamma_u(T)+\gamma_l(T)
\end{equation}
with $\gamma_u(T)$ ($\gamma_\ell(T)$) being the upper 
(lower) half width from the peak frequency, which is shown 
in Fig. \ref{F4_Peak_Gamma} (b). The saturation of the line width 
is due to the decay width $\Gamma_0$, which may originate from 
a phonon-electron interaction and a rattling phonon-acoustic 
phonon interaction neglected in this paper. 
However, at higher temperature, $\Gamma(T)$ increases fairly 
larger than $\Gamma_0$. Also the broadening in upper and lower 
frequency is asymmetric. The line broadening due to 
the anharmonicity leads to a non-Lorentzian spectral shape. 

From the analysis of this subsection, we see that 
anharmonicity of the rattling phonon leads to 
softening of the phonon frequency and sharpening and asymmetric 
change of the line width with decreasing temperature. 

\subsection{$k=0$}
\begin{figure}[h]
\includegraphics[width=6truecm]{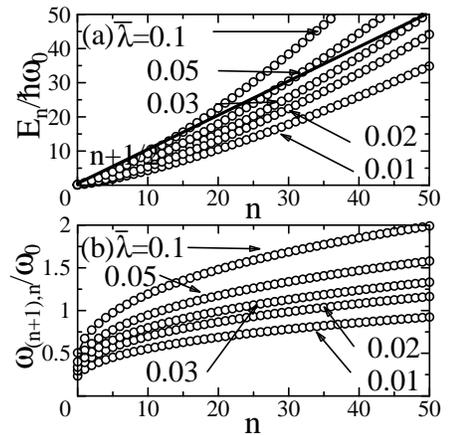}
\caption{(a) Energy eigenvalues and (b) excitation energy for $\bar k=0$}
\label{F5_eigenk0_w10} 
\end{figure}
\begin{figure}[h]
\includegraphics[width=6truecm]{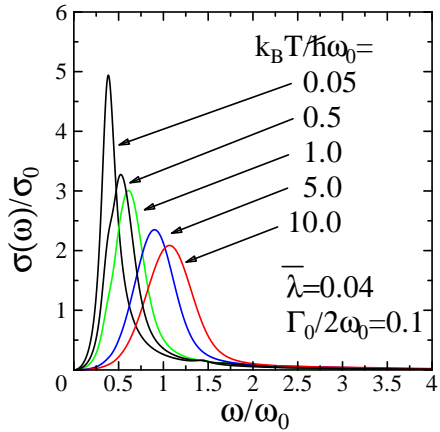}
\caption{(Color online) Temperature dependence of the optical conductivity 
for $\bar\lambda=0.04$ and $\bar k=0$.}
\label{F6_sigTk0l0_04} 
\end{figure}
\begin{figure}[h]
\includegraphics[width=6truecm]{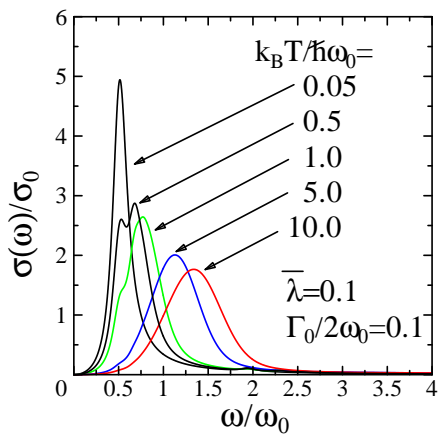}
\caption{(Color online) Temperature dependence of the optical conductivity 
for $\bar\lambda=0.1$ and $\bar k=0$.}
\label{F7_sigTk0l0_1} 
\end{figure}

When $k=0$ (or $|k|<<\lambda a_B^2$), 
the peak position and the line width 
are solely determined by $\lambda$. We choose $\hbar\omega_0$ as 
a characteristic energy of the system, for example, 
$\omega_0/2\pi$=1THz.  
In Figs. \ref{F5_eigenk0_w10} (a) and (b), 
the energy eigenvalues and the excitation energy 
$\omega_{(n+1),n}=(E_{n+1}-E_n)/\hbar$ are plotted. 
The lower excitation energies are softened and, just as well as 
in the case of $k>0$, the effectively contributing $n$ decreases 
with decreasing temperatures, resulting in the characteristic 
behavior, softening and sharpening, of the optical conductivity 
as is shown in Figs. \ref{F6_sigTk0l0_04}
and \ref{F7_sigTk0l0_1}. Note that the lowest excitation 
energy is softened up to about a half of the characteristic energy 
because of the shallow shape of the potential bottom.   

In the case of $\bar\lambda=0.04$ (Fig. \ref{F6_sigTk0l0_04}), 
the difference of the neighboring excitation energies is less 
than $\Gamma_0$, so that the optical conductivity shows 
a smooth softening, though the shift is enlarged and some 
shoulder is seen at low temperature ($k_BT/\hbar\omega_0=0.5$). 
Such a shoulder structure is smoothed for a smaller $\bar\lambda$.  
When $\bar\lambda$ becomes large, the difference of the lower 
excitation energies exceeds $\Gamma_0$, 
and as in Fig. \ref{F7_sigTk0l0_1} the optical 
conductivity shows a double peak structure as temperature 
decreased. The upper peak 
is reduced as temperature is further lowered. 
The broadening of the line width at higher temperature is 
enhanced as $\bar\lambda$ increases. 


\begin{figure*}[t]
\includegraphics[width=15truecm]{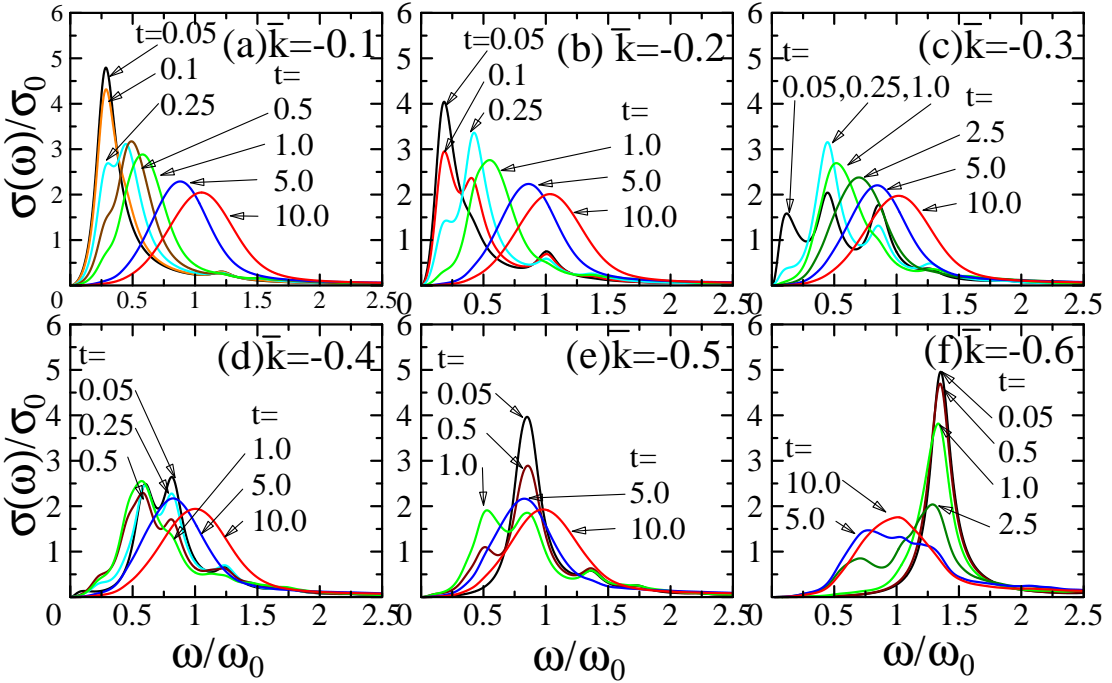}
\caption{(Color online, Wide figure) Temperature dependence of the optical conductivity 
for various $\bar k<0$. Parameters are$\bar\lambda=0.04$, 
$\Gamma_0/2\omega_0=0.1$. $t$ is the reduced 
temperature $t=k_BT/\hbar\omega_0$. }
\label{F8_sigTl0_04} 
\end{figure*}
\subsection{The case of $k<0$}
\begin{figure}[h]
\includegraphics[width=6truecm]{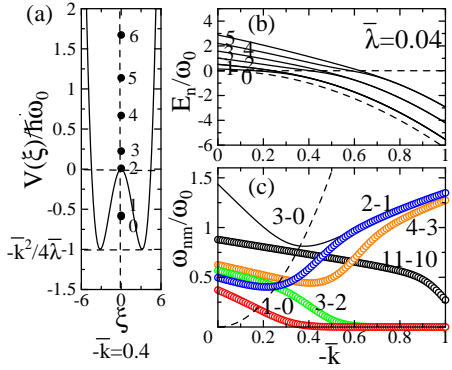}
\caption{(Color online) (a)Schematics of the double-minima potential and energy levels 
for $\bar k=-0.4$, and $\bar k$-dependence of 
(b)$E_{n}$ and (c)$\omega_{nm}$for 
low energy levels. $\bar\lambda$ is chosen as 0.04. }
\label{F9_PEnwnm} 
\end{figure}
For negative $k$, the anharmonic potential has double minima. 
We investigate the temperature 
dependence of the optical conductivity for $\bar\lambda=0.04$ 
by changing $\bar k$.  
Depending on the depth of the double minima, various patterns 
of temperature-dependent optical conductivity 
are obtained. We choose $\hbar\omega_0$ as 
a characteristic energy of the system, for example, 
$\omega_0/2\pi$=1THz. 

In Fig. \ref{F8_sigTl0_04}, we show the temperature dependence 
of the optical conductivity for various $\bar k$.  
The softening of the peak position is seen in 
Figs.\ \ref{F8_sigTl0_04}(a)-(c), but in Figs.\ \ref{F8_sigTl0_04}(d)-(f) 
the softening of the peak frequency is followed by 
the hardening, with decreasing temperature. 

Low temperature behaviors have various variety, 
especially there appear structures with double or triple peaks, though 
it depends on the magnitude of the width $\Gamma_0$. 
This can be understood from the low energy level transition in 
the double well potential. In Fig.\ \ref{F9_PEnwnm}(a), one example 
of the potential with $k<0$ is illustrated together with the 
eigenenergy levels. Low energy levels are much modified by the 
depth of the potential well. In Fig. \ref{F9_PEnwnm}(b) 
the $\bar k$-dependence of $E_{n}$ for 
low energy levels are plotted. Dotted lines are 
the potential double minima with the value 
$V(\pm \sqrt{-\bar k/\bar\lambda})/\hbar\omega_0
=-{\bar k}^2/4\bar\lambda$ and the local maximum $V(0)=0$, 
respectively. When $-\bar k$ increases, 
the levels $(E_0,E_1)$, $(E_2,E_3)$, $\cdots$, successively degenerate 
forming low-lying tunneling modes. 
In Fig. \ref{F9_PEnwnm}(c) the transition energies $\omega_{nm}$ are 
plotted among low-lying eigenstates. The dotted line is for the depth of 
the potential well. Transition energies 
$\omega_{10}$ and $\omega_{32}$ show successively 
softening as $-\bar k$ increases, while the neighboring 
excitations $\omega_{21}$, $\omega_{43}$ increase and 
become larger than excitation energies 
among higher levels ($\omega_{11,10}$ in Fig. \ref{F9_PEnwnm}(c)). 
The upturn of the excitation energies of $\omega_{21}$ is correlated 
with the depth of the potential well.    
In the following we discuss more details of low temperature 
behaviors in Fig.\ \ref{F8_sigTl0_04}. 

Figs. \ref{F8_sigTl0_04}(a) and (b) are 
for shallow double wells. 
Transition energy of $\hbar\omega_{10}$ is softened and  
it becomes smaller than $\hbar\omega_{21}$ 
due to the effect of the double well. Further, 
$(\omega_{21}-\omega_{10})$ is larger than 
$\Gamma_0$, so that there appears two 
peaks at low temperature and the higher 
peak of $\omega_{21}$ diminishes as $T\rightarrow 0$K. 
The peak around $\omega/\omega_0=1$ which remains even at $T=0$K 
corresponds to the transition $\hbar\omega_{30}$. 

Fig. \ref{F8_sigTl0_04}(c) is for $\bar k=-0.3$.  
The states 0 and 1 become very close but still have finite 
difference.   
We can identify lower peak as $\omega_{10}$, 
$\omega_{21}$ and $\omega_{30}$. 
The peak for $\omega_{21}$ is reduced with decreasing temperature 
(See also Fig. \ref{F9_PEnwnm}(c)).

Fig. \ref{F8_sigTl0_04}(d) is for $\bar k=-0.4$. 
The states 0 and 1 are almost degenerate, and this soft mode 
does not appear in the optical conductivity because of the factor 
$\omega$ in Eq. (\ref{sigEq}). 
The state 2 is inside the double well and the state 3 is 
above the maximum at $x=0$. Then $\omega_{21}$ is 
larger than $\omega_{32}$. We can identify the 
peaks in Fig. \ref{F8_sigTl0_04}(d) 
as $\omega_{32}$, $\omega_{12}$ and $\omega_{03}$ 
from the low energy side. Note that the intensity of the peak  
$\omega_{32}$ 
increases and decreases as temperature decreases. 

In Fig. \ref{F8_sigTl0_04}(e), 
$\bar k$ is further reduced 
as $\bar k=-0.5$. 
The state 2 and 3 start to degenerate and $\omega_{32}$ becomes 
lower than $\omega_{21}$.  We can identify the 
peaks in Fig. \ref{F8_sigTl0_04}(e) from 
the low energy side as $\omega_{43}$ and ($\omega_{12}$,$\omega_{03}$). 

In Fig. \ref{F8_sigTl0_04}(f),  
the state 2 and 3 are completely degenerate and  
$\omega_{21}$ is the main transition at low temperature.  
Since this energy is much larger than energies for higher 
level transition, the peak position first decrease at higher 
temperature and then increase at low temperature.  

When $\bar k$ is further reduced and the double minima become 
deep enough, the hardening of the peak frequency is obtained 
rather than softening as was discussed in ref. \cite{Foster1993}. 

In this way, we have various patterns depending on the strength 
of the double well

\subsection{Comparison with the experiment}
\begin{figure}[h]
\includegraphics[width=8truecm]{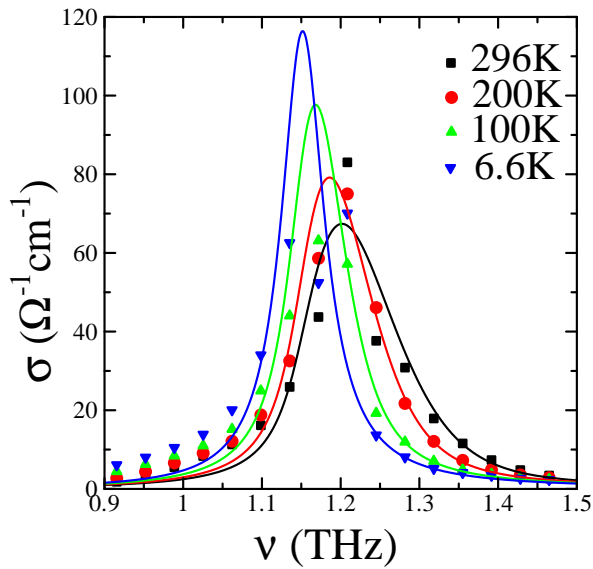}
\caption{(Color online) Comparison of the optical conductivity for 
Ba$_8$Ga$_{16}$Ge$_{30}$}
\label{F10_s1_RP} 
\end{figure}
	We have performed the time-domain THz spectroscopy and 
obtained the optical conductivity\cite{Mori2008b} in a type-I clathrate 
Ba$_8$Ga$_{16}$Ge$_{30}$\cite{Sales2001,Bentien2004,Avila2006,Madsen2005}. 
The details of the experimental method and analysis 
will be presented in a separated paper\cite{Mori2008b}. 
The obtained temperature-dependence of the 
phonon spectral for the lowest mode($\sim 1.2$THz) 
is shown by colored symbols in Fig. \ref{F10_s1_RP}. 
No multi-peak structure is observed. Then from the patterns presented 
in this paper, we conclude $k>0$ and we take $\bar k=1$.  

We adjust $\omega_0$, $\bar\lambda$, $\Gamma_0$ and  
$\sigma_0$ to fit overall behaviors and specially 
the higher frequency region (1.25$\sim$1.40 THz) of the spectral line. 
We choose  $\omega_0/2\pi=1.143$THz, $\bar\lambda=9.66\times 10^{-3}$, 
$\Gamma_0/2\pi=0.067$THz,    
and $\sigma_0=6.82\Omega^{-1}{\rm cm^{-1}}$. 
Theoretical results are presented by solid lines in Fig. \ref{F10_s1_RP}.   
The agreement between the experimental and theoretical results in 
temperature-dependence is very good.  

\begin{figure}[h]
\begin{minipage}{4truecm}
\includegraphics[width=4truecm]{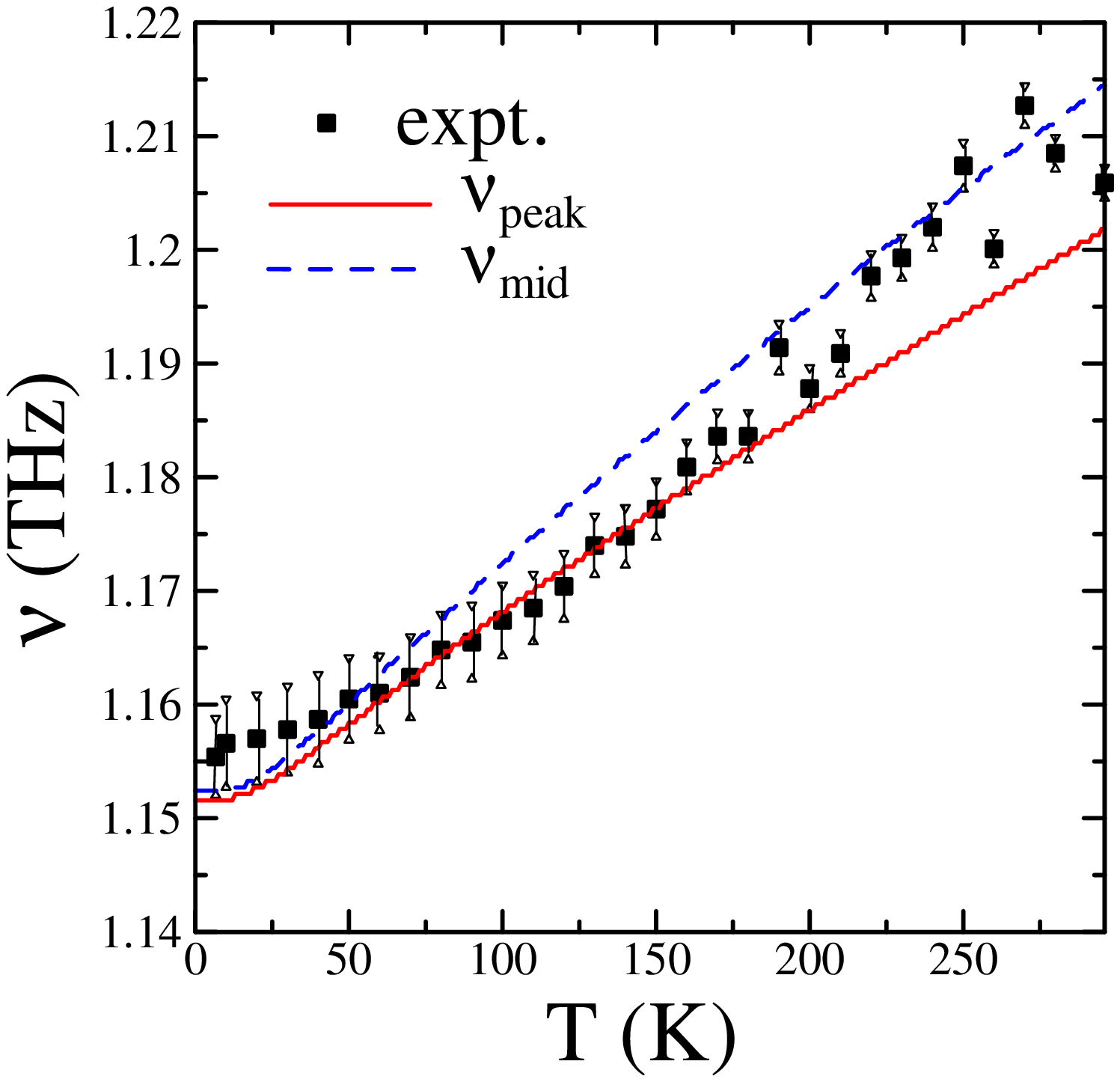}\\ 
(a) Peak frequency 
\end{minipage}
\begin{minipage}{4truecm}
\includegraphics[width=4truecm]{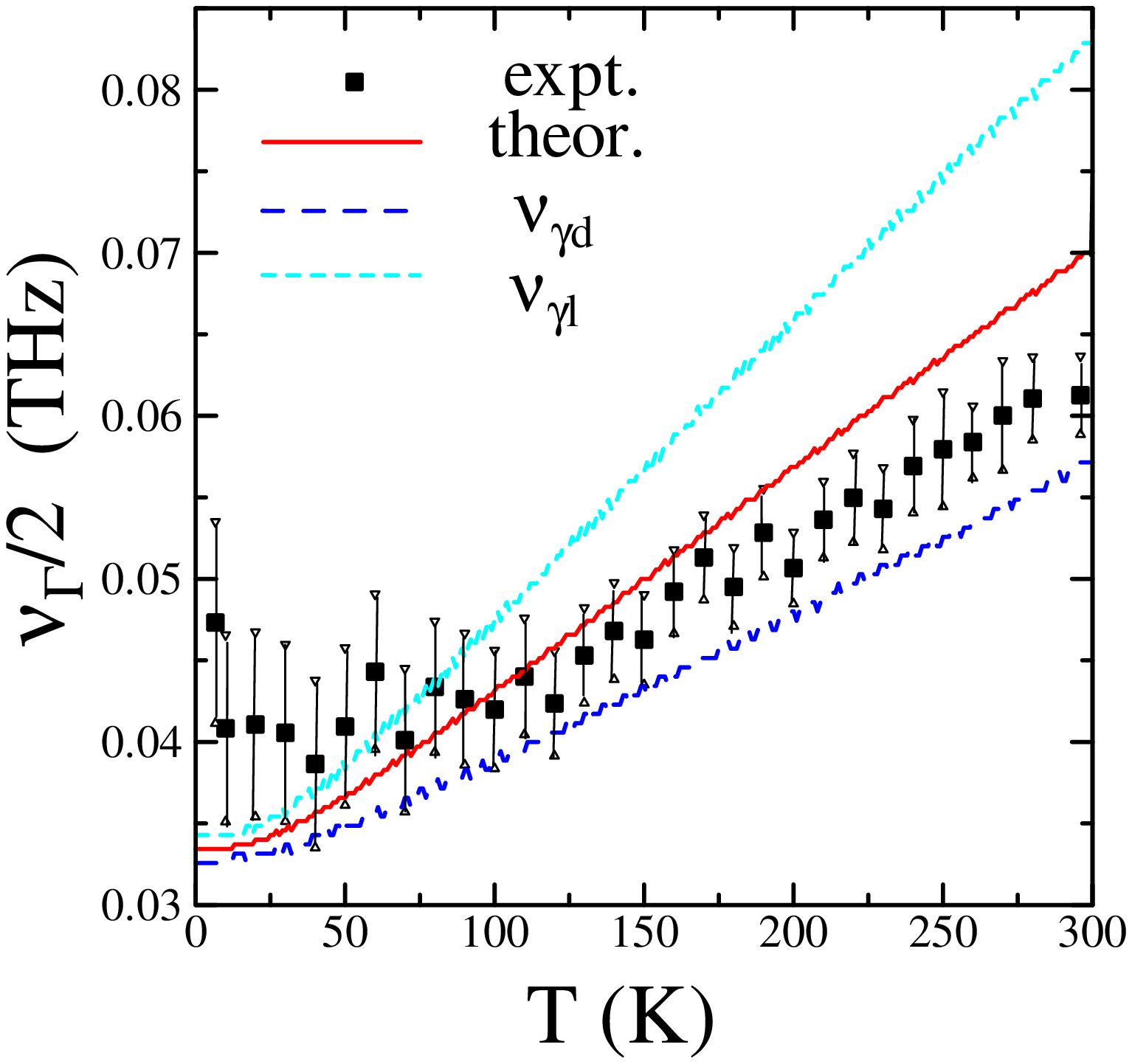}\\ 
(b) Line width\\  
\end{minipage}
\caption{(Color online) Temperature-dependence of the peak frequency and the line width}
				\label{F11_PWE} 
\end{figure}
The temperature dependence of the peak frequency and 
the line width are estimated from 
the experimental data by the 
Lorentzian fit and are plotted by the solid rectangulars in 
Fig. \ref{F11_PWE}(a) and (b), respectively. Theoretical 
results are plotted by solid and dashed lines. In Fig. \ref{F11_PWE}(a), 
the red solid line is for the peak frequency $\nu_{{\rm peak}}$ and the green 
dashed line is for the mid-frequency $\nu_{{\rm mid}}$. In Fig. \ref{F11_PWE}(b), 
the red solid, blue dashed and light blue short-dashed lines are 
frequencies $\nu_\Gamma/2$, 
$\nu_{\gamma_u}$ and $\nu_{\gamma_\ell}$ corresponding to 
the half width, the lower and upper half width, respectively. 
The experimental peak frequency situates 
between the theoretical peak and mid-frequency. Also the 
half-width is between the theoretical half-width and 
lower half-width. Since the Lorentzian fit is apt to lead 
a higher peak frequency and a narrower line width in an 
asymmetric line shape, the agreement between the experimental 
and theoretical results is reasonably good. 
The present comparison shows that the 
temperature-dependence of the rattling phonon 
in Ba$_8$Ga$_{16}$Ge$_{30}$ is well described 
by the anharmonicity of the potential without considering 
details of effects from other interactions. 
 
It has been reported that a type-I clathrate Ba$_8$Ga$_{16}$Sn$_{30}$ 
has an off-centered potential of the guest ion\cite{Avila2008}. 
It is an interesting problem to see if 
the optical conductivity of this material shows a behavior with $\bar k<0$ 
discussed in this paper\cite{note2}.   

\section{Conclusion}
In this paper we have investigated theoretically 
the temperature dependence 
of the optical conductivity from the rattling phonon. The guest 
ion feels an anharmonic potential from the cage, and the anharmonic 
effect appears characteristically in the temperature dependence, 
that is, the softening of the peak frequency and sharpening of 
the line width with decreasing temperature. 
 
In the case of the positive quadratic term, one can expect 
the quadratic coefficient is roughly determined from the 
saturated peak frequency at low temperature and the quartic term is 
determined from the line width and shift of the peak frequency. 

In the case of the negative quadratic term, various patterns 
of the optical conductivity are expected depending on the 
strength of the double minima in the potential. 
Multi-peak structures appear at low 
temperature and increase and decrease of the peak frequency  
with temperature are obtained. 
 
We have shown that measurements on the 
temperature dependence of the optical conductivity can 
provide a direct evidence for the anharmonicity.

\section*{\bf Acknowledgements} 
We thank Drs. M. A. Avila, J. P. Carbotte, M. Dressel, H. Hasegawa, M. Lang, H.
Matsui, T. Nakayama, N. Ogita, K. Suekuni, T. Takabatake, Y. Takasu, K.
Tanigaki, M. Udagawa, K. Ueda, W. Weber, A. Yamakage, A. Yoshihara, M.
Yoshizawa and T. Koyama for valuable discussions. Two of us (T. M., K. I.) are 
supportedfinancially by the Global COE program ``Materials Integrations", Tohoku
University. This work has been supported by Grants-in-Aid for Scientific
Research (A)(15201019) and the priority area ``Nanospace'' from MEXT,
Japan.

\begin{thebibliography}{99}
\bibitem{note1} In ref. \cite{Elliot1965}, 
	the hardening of the local mode 
	with decreasing temperature was indicated because of the effect 
	from the band phonon, and in ref. \cite{Foster1993}, a deep double 
    minumum potential was considered, leading also the hardening of the 
    local mode with decreasing temperature."
\end{thebibliography}

\end{document}